# On Optimal Distributed Joint Source-Channel Coding for Correlated Gaussian Sources over Gaussian Channels

R Rajesh and Vinod Sharma *Senior Member IEEE*

**Abstract**

We consider the problem of distributed joint source-channel coding of correlated Gaussian sources over a Gaussian Multiple Access Channel (GMAC). There may be side information at the decoder and/or at the encoders. First we specialize a general result in [20] to obtain sufficient conditions for reliable transmission over a Gaussian MAC. This system does not satisfy the source-channel separation. We study and compare three joint source-channel coding schemes available in literature. We show that each of these schemes is optimal under different scenarios. One of the schemes, Amplify and Forward (AF) which simplifies the design of encoders and the decoder, is optimal at low SNR but not at high SNR. Another scheme is asymptotically optimal at high SNR. The third coding scheme is optimal for orthogonal Gaussian channels. We also show that AF is close to the optimal scheme for orthogonal channels even at high SNR.

Keywords: Gaussian multiple access channel, side information, separation-based transmission, amplify and forward, correlated sources, orthogonal channels.

## I. INTRODUCTION AND SURVEY

Sensor networks are used in a wide variety of applications, the most common being the spatio-temporal monitoring of a random field ([1]) and the detection of change in its statistics ([23]). Sensor nodes are inexpensive with limited battery power and storage and hence have limited computing and communication capabilities ([1]). These nodes transmit their observations to a fusion center to estimate the sensed random field. Since transmission is very energy intensive, it is important to minimize it.

The sensor nodes transmit their observations to the fusion center (or a cluster head) usually over a Multiple Access Channel (MAC) ([2], [23]). Often the received symbol is a super-position of the transmitted symbols corrupted by Additive White Gaussian Noise (AWGN). This then is the well known Gaussian MAC (GMAC). This channel is interesting from a practical as well as a theoretical perspective.

Also the sensor nodes can be modeled as discrete or continuous sources. For continuous sources, Gaussian distribution is particularly useful. This for example can happen if the sensor nodes are sampling a Gaussian random field. We can come across it in the problem of detection of change also. Then, it is often the detection of change in the mean of the sensor observations with the sensor observation noise being Gaussian ([23]). Thus, in this paper we focus on transmission of Gaussian sources over a GMAC.

One common way to use the MAC is via Time division multiple access (TDMA), Frequency division multiple access (FDMA) or Code division multiple access (CDMA) ([3], [7]). These protocols although suboptimal are used due to practical considerations. These protocols make the MAC a set of parallel orthogonal channels (for CDMA, it happens if we use orthogonal codes). Thus we will also consider orthogonal Gaussian channels.

In the following we survey the related literature. Cover, El Gamal and Salehi [6] provided sufficient conditions for transmitting losslessly discrete correlated observations over a discrete MAC and show that source-channel separation does not hold for this system. The results of [6] have been extended in [20] to the case of lossy transmission with side information and continuous alphabets.

The distributed Gaussian source coding problem is discussed in [15], [24]. The exact rate region for two users is provided in [24]. In [14] one necessary and two sufficient conditions for transmitting a bivariate jointly Gaussian source over a GMAC are provided. The authors prove that the (uncoded) amplify and forward (AF) scheme is optimal below a certain SNR.

In [17] it is shown that feedback increases the capacity of a GMAC. GMAC with lossy transmission of correlated discrete sources and side information is studied in [19]. Transmission of correlated jointly Gaussian input over a GMAC is also studied in [18] and the edge capacities obtained are similar to the expressions in [17]. In [10] the authors discuss a joint source channel coding scheme over a MAC and show the scaling behavior for the Gaussian channel. Actually their problem is closer to the Gaussian CEO problem [16]. The scaling laws for the problem without side information are discussed in [12] and it is shown that separating source coding from channel coding may require exponential increase in bandwidth, as the number of sensors increases. In [13], the authors show that for a Gaussian sensor network it is better to compress the local estimates than to compress the raw data.

Conditions for separation to hold in multiple access channels are given in [21]. Separation holds for a GMAC under receiver power constraints ([9]). In general, for orthogonal channels separation holds for lossless ([3]) as well as for lossy transmission ([26]). It is shown in [11] that separation holds and uncoded transmission achieves capacity in a Gaussian relay network as the number of relays go to infinity.

This paper makes the following contributions. From our general results in [20] we obtain explicit conditions for transmission of correlated Gaussian sources with given distortion over a GMAC. Also we compare the two schemes in [14] with a separation based scheme. We explicitly show that the AF scheme in [14] is not optimal at high SNR. However another scheme studied in [14] is. Furthermore, for AF, it may not be optimal to use all the power. We provide the results with side information also. We identify an optimal coding scheme for transmission of correlated Gaussian sources over orthogonal Gaussian channels and compare it to AF. We show that AF is close to the optimal scheme. We also compare the performance with side information. The results are extended to more than two users providing some unexpected conclusions.

The paper is organized as follows. Sufficient conditions for transmission of continuous correlated sources over a continuous MAC are given in Section II. The transmission of Gaussian sources on a Gaussian MAC is discussed in Section III. Different joint source-channel coding schemes for transmission are studied and their asymptotic performances are compared. In Section IV optimal power allocation to minimize the sum of the distortions for the three schemes is obtained. Section V gives the performance with side information. In Section VI transmission of correlated Gaussian sources over orthogonal Gaussian channels is considered. The performance of the AF scheme is studied VII. Section VIII concludes the paper.

## II. Transmission of correlated sources over a MAC

In this section we consider the transmission of memoryless dependent sources, through a memoryless multiple access channel. The sources and/or the channel input/output alphabets can be discrete or continuous. Furthermore, side information about the transmitted information may be available at the encoders and the decoder. Thus our system is very general and covers many systems studied earlier.

We consider two sources $(U_1, U_2)$ and side information random variables $Z_1, Z_2, Z$ with a known joint distribution $F(u_1, u_2, z_1, z_2, z)$ (generalization to multiple sources is also available). Side information $Z_i$ is available to encoder $i$, $i = 1, 2$ and the decoder has side information $Z$. The random vector sequence $\{(U_{1n}, U_{2n}, Z_{1n}, Z_{2n}, Z_n), n \geq 1\}$ formed from the source outputs and the side information with distribution $F$ is independent identically distributed (*iid*) in time. We will denote $\{U_{1k}, \ k = 1, ..., n\}$ by $U_1^n$. Similarly for other sequences. The sources transmit their codewords $X_{in}$'s to a single decoder through a memoryless multiple access channel. The channel output $Y$ has distribution $p(y|x_1, x_2)$ if $x_1$ and $x_2$ are transmitted at that time. Thus, $\{Y_n\}$ and $\{(X_{1n}, X_{2n})\}$ satisfy $p(y_k|y^{k-1}, x_1^k, x_2^k) = p(y_k|x_{1k}, x_{2k})$. The decoder receives $Y_n$ and also has access to the side information $Z_n$. The encoders at the two users do not communicate with each other except via the side information. The decoder uses the channel outputs and its side information to estimate the sensor observations $U_{in}$ as $\hat{U}_{in}$, $i = 1, 2$. It is of interest to find encoders and a decoder such that $\{U_{1n}, U_{2n}, n \geq 1\}$ can be transmitted over the given MAC with $E[d_1(U_1, \hat{U}_1)] \leq D_1$ and $E[d_2(U_2, \hat{U}_2)] \leq D_2$ where $d_i$ are non-negative distortion measures and $D_i$ are the given distortion constraints. If the distortion measures are unbounded we assume that $u_i^*$, $i \in \{1, 2\}$ exist such that $E[d_i(U_i, u_i^*)] < \infty$, $i = 1, 2$. Source channel separation does not hold in this case.

*Definition*: The source $(U_1^n, U_2^n)$ can be transmitted over the multiple access channel with distortions $\mathbf{D} \triangleq (D_1, D_2)$ if for any $\epsilon > 0$ there is an $n_0$ such that for all $n > n_0$ there exist encoders $f_{E,i}^n : \mathcal{U}_i^n \times \mathcal{Z}_i^n \to \mathcal{X}_i^n$, $i = 1, 2$ and a decoder $f_D^n : \mathcal{Y}^n \times \mathcal{Z}^n \to (\hat{\mathcal{U}}_1^n, \hat{\mathcal{U}}_2^n)$ such that $\frac{1}{n} E \left[ \sum_{j=1}^n d(U_{ij}, \hat{U}_{ij}) \right] \leq D_i + \epsilon$, $i = 1, 2$ where $(\hat{U}_1^n, \hat{U}_2^n) = f_D(Y^n, Z^n)$ and $\mathcal{U}_i$, $\mathcal{Z}_i$, $\mathcal{Z}$, $\mathcal{X}_i$, $\mathcal{Y}$, $\hat{\mathcal{U}}_i$ are the sets in which $U_i$, $Z_i$, $Z$, $X_i$, $Y$, $\hat{U}_i$ take values.

We denote the joint distribution of $(U_1, U_2)$ by $p(u_1, u_2)$. $X \leftrightarrow Y \leftrightarrow Z$ will indicate that $\{X, Y, Z\}$ form a Markov chain.

The proof of the following theorem is available in [20] for more than two users also. It will be specialized to the Gaussian sources and GMAC and then used to obtain efficient joint source-channel coding schemes.

*Theorem 1:* A source $(U_1, U_2)$ can be transmitted over the multiple access channel with distortions $(D_1, D_2)$ if there exist random variables $(W_1, W_2, X_1, X_2)$ such that

(1) $p(u_1, u_2, z_1, z_2, z, w_1, w_2, x_1, x_2, y) = p(u_1, u_2, z_1, z_2, z) p(w_1 | u_1, z_1) p(w_2 | u_2, z_2).$

$$p(x_1 | w_1) p(x_2 | w_2) p(y | x_1, x_2).$$

(2) There exists a function $f_D : \mathcal{W}_1 \times \mathcal{W}_2 \times \mathcal{Z} \to (\hat{\mathcal{U}}_1 \times \hat{\mathcal{U}}_2)$ such that $E[d(U_i, \hat{U}_i)] \leq D_i$, $i = 1, 2$, where $(\hat{U}_1, \hat{U}_2) = f_D(W_1, W_2, Z)$, $\mathcal{W}_i$ are the sets in which $W_i$ take values and the constraints

$$\begin{aligned} I(U_1, Z_1; W_1 | W_2, Z) &< I(X_1; Y | X_2, W_2, Z), \\ I(U_2, Z_2; W_2 | W_1, Z) &< I(X_2; Y | X_1, W_1, Z), \\ I(U_1, U_2, Z_1, Z_2; W_1, W_2 | Z) &< I(X_1, X_2; Y | Z), \end{aligned} \quad (1)$$

are satisfied. ∎

If the channel alphabets are continuous (e.g., GMAC) then in addition to the conditions in Theorem 1 certain power constraints $E[X_i^2] \leq P_i$, $i = 1, 2$ are also needed. In general, we could impose a constraint $E[g_i(X_i)] \leq \alpha_i$ where $g_i$ is some non-negative cost function. Furthermore, for continuous alphabet r.v.s ( sources/channel input/output) we will assume that probability density exists so that one can use differential entropy.

The correlations in $(U_1, U_2)$ and the side information $(Z_1, Z_2, Z)$ are used to decrease the left side and increase the right side in (1).

We specialize this result to the Gaussian sources and GMAC in the next section. The main problem in applying the result in the above theorem in specific examples is to obtain a good coding scheme $(W_1, W_2, X_1, X_2)$ (as is usual in most information theoretic results). Thus we will also consider specific coding schemes and study their performance.

In the first part of this paper we will mostly consider the system without side information $Z_1$, $Z_2$, $Z$. Then the inequalities in (1) become

$$I(U_1; W_1|W_2) < I(X_1; Y|X_2, W_2), \ I(U_2; W_2|W_1) < I(X_2; Y|X_1, W_1),$$
$$I(U_1, U_2; W_1, W_2) < I(X_1, X_2; Y). \quad (2)$$

It has been shown in [19] that under our conditions $I(X_1; Y|X_2, W_2) \leq I(X_1; Y|X_2)$ and $I(X_2; Y|X_1, W_1) \leq I(X_2; Y|X_1)$. We will use these relaxed upper bounds to obtain good joint source-channel coding schemes.

III. GAUSSIAN SOURCES OVER GAUSSIAN MAC

In a Gaussian MAC the channel output $Y_n$ at time $n$ is given by $Y_n = X_{1n} + X_{2n} + N_n$ where $X_{1n}$ and $X_{2n}$ are the channel inputs at time $n$ and $N_n$ is a Gaussian random variable independent of $X_{1n}$ and $X_{2n}$, with $E[N_n] = 0$ and $var(N_n) = \sigma_N^2$ (we will denote this distribution by $\mathcal{N}(0, \sigma_N^2)$). We will also assume that $(U_{1n}, U_{2n})$ is jointly Gaussian with mean zero, variances $\sigma_i^2$, $i = 1, 2$ and correlation $\rho$. The distortion measure will be Mean Square Error (MSE). The transmission power constraints are $E[X_i^2] \leq P_i$, $i = 1, 2$.

It is shown in [20] that $I(X_1; Y|X_2)$, $I(X_2; Y|X_1)$ and $I(X_1, X_2; Y)$ are maximized by zero mean Gaussian random variables $(X_1, X_2)$ with variances $P_1$ and $P_2$. Also if correlation between $X_1$ and $X_2$

is $\tilde{\rho}$ then the conditions in (2) (after the relaxation mentioned below (2)) become

$$I(U_1;W_1|W_2) < 0.5\log\left[1+\frac{P_1(1-\tilde{\rho}^2)}{\sigma_N{}^2}\right], \ I(U_2;W_2|W_1) < 0.5\log\left[1+\frac{P_2(1-\tilde{\rho}^2)}{\sigma_N{}^2}\right],$$
$$I(U_1,U_2;W_1,W_2) < 0.5\log\left[1+\frac{P_1+P_2+2\tilde{\rho}\sqrt{P_1P_2}}{\sigma_N{}^2}\right]. \tag{3}$$

An advantage of the relaxed conditions (3) is that we will be able to obtain a good source-channel coding scheme. Once $(X_1, X_2)$ are obtained we can check for sufficient conditions (2). If these conditions are not satisfied then we increase $\tilde{\rho}$ till (2) is satisfied. See more details in [19].

In the rest of the paper we consider three specific coding schemes to obtain $W_1, W_2, X_1, X_2$ where $(W_1, W_2)$ satisfy the distortion constraints in Theorem 1 and $(X_1, X_2)$ are jointly Gaussian with an appropriate $\tilde{\rho}$ such that (3) (and then (2)) is satisfied. These coding schemes have been available before. Our purpose is to compare their performance. More importantly we show that each of these coding schemes can be optimal under different scenarios. Also in two of these coding schemes $X_i$ is obtained from $W_i$ by scaling. Thus, the relaxed conditions (3) in fact are identical to the sufficient conditions obtained in Theorem 1. In the third coding scheme (SB), $W_1$ and $W_2$ are (asymptotically) independent due to Slepian-Wolf coding. Thus again $I(X_1; Y|X_2, W_2) = I(X_1; Y|X_2)$ and $I(X_2; Y|X_1, W_1) = I(X_2; Y|X_1)$.

*A. Amplify and forward scheme*

In the Amplify and Forward (AF) scheme the channel codes $X_i$ are just scaled source symbols $U_i$. Since $(U_1, U_2)$ are themselves jointly Gaussian, $(X_1, X_2)$ will be jointly Gaussian and retain the dependence of inputs $(U_1, U_2)$. The scaling is done to ensure $E[X_i{}^2] = P_i, i = 1, 2$. For a single user case this coding is optimal ([7]).

At the decoder inputs $U_1$ and $U_2$ are directly estimated from $Y$ as $\hat{U}_i = E[U_i|Y], \ i = 1, 2$. Because $U_i$ and $Y$ are jointly Gaussian this estimate is linear and also satisfies the Minimum Mean Square Error (MMSE) and the Maximum Likelihood (ML) criteria.

The MMSE distortion for this encoding-decoding scheme can be easily shown to be

$$\overline{D_1} = \frac{\sigma_1{}^2\left[P_2(1-\rho^2)+\sigma_N{}^2\right]}{P_1+P_2+2\rho\sqrt{P_1P_2}+\sigma_N{}^2}, \ \overline{D_2} = \frac{\sigma_2{}^2\left[P_1(1-\rho^2)+\sigma_N{}^2\right]}{P_1+P_2+2\rho\sqrt{P_1P_2}+\sigma_N{}^2}. \tag{4}$$

Since encoding and decoding require minimum processing and delay in this scheme, if it satisfies the required distortion bounds $D_i$, it should be the scheme to implement. This scheme has been studied in [14] and found to be optimal below a certain SNR for two-user symmetric case ($P_1 = P_2, \sigma_1 = \sigma_2, D_1 = D_2$). However unlike for single user case, in this case user 1 acts as interference for user 2 (and vice versa).

Thus one should not expect this scheme to be optimal under high SNR case. Also, in the asymmetric case, it is not clear if all the power $P_1$, $P_2$ needs to be expended for AF transmission. We will address these issues in the following sections.

*B. Separation based scheme*

In separation based (SB) approach the jointly Gaussian sources are vector quantized to $W_1^n$ and $W_2^n$. The quantized outputs are Slepian-Wolf encoded [22]. This produces code words, which are (asymptotically) independent. The rate $(R_1, R_2)$, distortion $(D_1, D_2)$ pairs achievable for this source coding are ([24])

$$\begin{aligned} R_1 &\geq 0.5 \log\left[\sigma_1^2(1 - \rho^2 + \rho^2 2^{-2R_2})/D_1\right], \\ R_2 &\geq 0.5 \log\left[\sigma_2^2(1 - \rho^2 + \rho^2 2^{-2R_1})/D_2\right], \\ R_1 + R_2 &\geq 0.5 \log\left[\sigma_1^2 \sigma_2^2 (1 - \rho^2)\beta(D_1, D_2)/2D_1 D_2\right], \end{aligned} \qquad (5)$$

where $\beta(D_1, D_2) = 1 + \sqrt{1 + 4\rho^2 D_1 D_2 / \sigma_1^2 \sigma_2^2 (1 - \rho^2)^2}$.

The independent code words $(W_1^n, W_2^n)$ are encoded to capacity achieving independent Gaussian channel codes $(X_1^n, X_2^n)$. This is a very natural scheme and has been considered by various authors ([6], [7], [21]).

Since source-channel separation does not hold for this system, this scheme is not expected to be optimal (however we will see that it is optimal for orthogonal channels). But because this scheme decouples source coding from channel coding, it is preferable to a joint source-channel coding scheme with comparable performance.

*C. Lapidoth-Tinguely scheme*

In this scheme, obtained in [14], $(U_1^n, U_2^n)$ are vector quantized to $2^{nR_1}$, $2^{nR_2}$ $(\tilde{U}_1^n, \tilde{U}_2^n)$ vectors where $R_1$ and $R_2$ will be specified below. Also, $W_1^n, W_2^n$ are $2^{nR_1}$ and $2^{nR_2}$, $n$ length code words obtained independently with distributions $\mathcal{N}(0, 1)$. For each $\tilde{u}_i^n$, we pick the codeword $w_i^n$ that is closest to it. This way we obtain Gaussian codewords $W_1^n, W_2^n$ which retain the correlations of $(U_1^n, U_2^n)$. $X_1^n$ and $X_2^n$ are obtained by scaling $W_1^n, W_2^n$ to satisfy the transmit power constraints. Thus (3) is a *sufficient* condition for this scheme. We will call this scheme LT. $(U_1, U_2, W_1, W_2)$ are (approximately) jointly Gaussian with covariance matrix

$$\begin{pmatrix} \sigma_1^2 & \rho\sigma_1\sigma_2 & \sigma_1^2(1 - 2^{-2R_1}) & \rho\sigma_1\sigma_2(1 - 2^{-2R_2}) \\ \rho\sigma_1\sigma_2 & \sigma_2^2 & \rho\sigma_1\sigma_2(1 - 2^{-2R_1}) & \sigma_2^2(1 - 2^{-2R_2}) \\ \sigma_1^2(1 - 2^{-2R_1}) & \rho\sigma_1\sigma_2(1 - 2^{-2R_1}) & \sigma_1^2(1 - 2^{-2R_1}) & \frac{\tilde{\rho}^2 \sigma_1 \sigma_2}{\rho} \\ \rho\sigma_1\sigma_2(1 - 2^{-2R_2}) & \sigma_2^2(1 - 2^{-2R_2}) & \frac{\tilde{\rho}^2 \sigma_1 \sigma_2}{\rho} & \sigma_2^2(1 - 2^{-2R_2}) \end{pmatrix}, \qquad (6)$$

where $\tilde{\rho} = \rho\sqrt{(1-2^{-2R_1})(1-2^{-2R_2})}$. The required $(R_1, R_2)$ are obtained from (3) as follows. From $I(U_1; W_1|W_2) = H(W_1|W_2) - H(W_1|W_2, U_1)$ and the fact that the Markov chain condition $W_1 \leftrightarrow U_1 \leftrightarrow U_2 \leftrightarrow W_2$ holds, we obtain $H(W_1|W_2, U_1) = H(W_1|U_1)$ and $I(U_1; W_1|W_2) = 0.5 \log \left[(1-\tilde{\rho}^2)2^{2R_1}\right]$.

Thus from (3) (which is now the same as (2) because $X_i$ are obtained from $W_i$ via scaling) we need $R_1$ and $R_2$ which satisfy

$$R_1 \leq 0.5 \log \left[\frac{P_1}{\sigma_N^2} + \frac{1}{(1-\tilde{\rho}^2)}\right], \quad R_2 \leq 0.5 \log \left[\frac{P_2}{\sigma_N^2} + \frac{1}{(1-\tilde{\rho}^2)}\right],$$
$$R_1 + R_2 \leq 0.5 \log \left[\frac{\sigma_N^2 + P_1 + P_2 + 2\tilde{\rho}\sqrt{P_1 P_2}}{(1-\tilde{\rho}^2)\sigma_N^2}\right]. \tag{7}$$

The inequalities (7) are the same as in [14]. Thus we recover the conditions in [14] from our general result (1).

Taking $\hat{U}_i = E[U_i|W_1, W_2]$, $i = 1, 2$, we obtain the distortions

$$D_1 = var(U_1|W_1, W_2) = \frac{\sigma_1^2 2^{-2R_1}\left[1 - \rho^2\left(1 - 2^{-2R_2}\right)\right]}{(1-\tilde{\rho}^2)}, \tag{8}$$

$$D_2 = var(U_2|W_1, W_2) = \frac{\sigma_2^2 2^{-2R_2}\left[1 - \rho^2\left(1 - 2^{-2R_1}\right)\right]}{(1-\tilde{\rho}^2)}. \tag{9}$$

The minimum distortion is obtained when $\tilde{\rho}$ is such that the sum rate is met with equality in (7).

### D. Asymptotic performance of the three schemes

We compare the performance of the three schemes. We will extend the comparison to more than two users in Section III-E. For simplicity we consider the symmetric case: $P_1 = P_2 = P$, $\sigma_1 = \sigma_2 = \sigma$, $D_1 = D_2 = D$. We will denote the SNR $P/\sigma_N^2$ by $S$ and $D(S)$ will denote the distortion for a given $S$.

Consider the AF scheme. From (4),

$$D(S) = \frac{\sigma^2 \left[S\left(1 - \rho^2\right) + 1\right]}{2S\left(1 + \rho\right) + 1}. \tag{10}$$

Thus $D(S)$ decreases to $\sigma^2(1-\rho)/2$ strictly monotonically at rate $O(1)$ as $S \to \infty$. Also,

$$\lim_{S \to 0} \left|\frac{D(S) - \sigma^2}{S}\right| = \sigma^2(1+\rho)^2. \tag{11}$$

Hence, $D(S) \to \sigma^2$ at rate $O(S)$ as $S \to 0$.

Next consider the SB scheme. From [24] if each source is encoded at rate $R$, it can be decoded at the decoder with distortion

$$D = \sqrt{2^{-4R}(1-\rho^2) + \rho^2 2^{-8R}}. \tag{12}$$

At high SNR, from the capacity result for independent inputs, we have $R < 0.25 \log S$ ([7]). Then from (12) we obtain

$$D \geq \sqrt{\frac{\sigma^4(1-\rho^2)}{S} + \frac{\sigma^4\rho^2}{S^2}} \tag{13}$$

and this lower bound is achievable. As $S \to \infty$, this lower bound approaches zero at rate $O(\sqrt{S})$. Thus this scheme outperforms AF at high SNR. At low SNR, $R \approx S/2$ and hence from (12)

$$D \geq \rho^2\sigma^4 2^{-4S} + \sigma^2(1-\rho^2)2^{-2S}. \tag{14}$$

Thus $D \to \sigma^2$ at rate $O(S^2)$ as $S \to 0$ at high $\rho$ and at rate $O(S)$ at small $\rho$. Therefore we expect that at low SNR, at high $\rho$ this scheme will be worse than AF but at low $\rho$ it will be comparable.

Consider the LT scheme. In the high SNR region we assume that $\tilde{\rho} = \rho$ since $R = R_1 = R_2$ are sufficiently large. Then from (7) $R \approx 0.25 \log[2S/(1-\rho)]$ and the distortion can be approximated by (and asymptotically approaches)

$$D \approx \sigma^2 \sqrt{(1-\rho)/2S}. \tag{15}$$

Therefore, $D \to 0$ as $S \to \infty$ at rate $O(\sqrt{S})$. This rate of convergence is same as for SB. However, the right side in (13) is greater than that of (15) and at low $\rho$ the two are close. Thus at high SNR LT always outperforms SB but the improvement is small for low $\rho$. Now we show that LT is in fact asymptotically optimal at high SNR.

The necessary conditions (NC) to be able to transmit on the GMAC with distortion $(D, D)$ for the symmetric case are [14], [25]

$$D \geq \begin{cases} \frac{\sigma^2[S(1-\rho^2)+1]}{2S(1+\rho)+1}, & \text{for } S \leq \frac{\rho}{1-\rho^2}, \\ \sigma^2\sqrt{\frac{(1-\rho^2)}{2S(1+\rho)+1}}, & \text{for } S > \frac{\rho}{1-\rho^2}. \end{cases} \tag{16}$$

At high SNR the lower bound tends to $\sigma^2\sqrt{(1-\rho)/2S}$. This is same as the expression for LT in (15).

At low SNR

$$R \approx \frac{S(1+\tilde{\rho})}{2} - \frac{\log(1-\tilde{\rho}^2)}{4}$$

and evaluating $D$ from (8) we get

$$D = \frac{\sigma^2 2^{-\bar{S}}\left(1 - \rho^2(1 - \sqrt{1-\tilde{\rho}^2}2^{-\bar{S}})\right)}{\sqrt{1-\tilde{\rho}^2}} \tag{17}$$

where $\overline{S} = S(1+\tilde{\rho})$. Therefore $D \to \sigma^2$ as $S \to 0$ at rate $O(S^2)$ at high $\rho$ and at rate $O(S)$ at low $\rho$. These rates are the same as that for SB. In fact, dividing the expression for $D$ at low SNR for SB by that for LT, we can show that the two distortions tend to $\sigma^2$ at the same rate for all $\rho$.

The above three schemes along with the necessary conditions are compared below using exact computations. Figures 1 and 2 show the distortion as a function of SNR for unit variance jointly Gaussian sources with correlations $\rho = 0.1$ and $0.75$.

From these plots we confirm our theoretical conclusions provided above. In particular we obtain that:

- AF is close to necessary conditions and hence optimal (as shown in [14]) at low SNR. The other two schemes perform worse at low SNR.
- SB and LT perform better than AF at high SNRs.
- LT performs better than SB in general.
- Performance of SB and LT are close for low $\rho$ and for high $\rho$ at low SNR.
- LT is close to optimal at high SNR and is asymptotically optimal.

### E. Multiple users

In this section we study the performance of the three schemes for more than two users. This is important for sensor networks. We limit ourselves to symmetric scenario (this is for convenience, the general case can be handled similarly). Consider jointly Gaussian $(U_1, U_2, ..., U_N)$ having mean zero vector with $E(U_i, U_j) = \rho$, $i \neq j$ and $E[U_i^2] = 1$. Let all the nodes be (average) power constrained to $P$.

Consider the AF scheme. Let $D_{AF}(N,P)$ denote the distortion per user incurred in transmitting the sources through a GMAC. For this scheme then

$$D_{AF}(N,P) = 1 - \frac{P(1+(N-1)\rho)^2}{NP(1+(N-1)\rho) + \sigma^2}. \tag{18}$$

Also,

$$\lim_{N \to \infty} D_{AF}(N,P) = 1 - \rho, \quad \lim_{P \to \infty} D_{AF}(N,P) = \frac{N-1}{N}(1-\rho). \tag{19}$$

From (19) we can see that distortion per user tends to $1 - \rho$ for all $P$. Thus correlation reduces the asymptotic mean distortion. Also from (19) we see that, at high SNR, distortion per user increases from $(1-\rho)/2$ for the two user case to $1 - \rho$ as $N \to \infty$.

We plot $D_{AF}(N,P)$ for $\rho = 0.8$ in Fig. 3. Interestingly, we find that for low SNR as $N$ increases $D_{AF}(N,P)$ decreases. But at high SNR as $N$ increases $D_{AF}(N,P)$ increases. This happens because of the trade-off between the interference caused by many users and the beamforming gain (because of the correlation in the observations). This trade-off depends on the SNR. The cut-off where this change (decrease to increase) occurs depends on $\rho$. The cut-off increases with $\rho$. For $\rho = 0.8$ and for $N \leq 10$ the cutoff could be taken as $3dB$. Thus, for $N \leq 10$, for $P \leq 3dB$ distortion $D(N, P)$ decreases with $N$.

Consider the SB scheme. In this coding scheme the source output $U_i^n \triangleq (U_{i1}, ..., U_{in})$, $i = 1, ...N$ is vector quantized to $W_i^n$ by sensor node $S_i$. The $W_i^n$'s are Slepian-Wolf encoded. The encoded data is sent over the channel using independent channel codewords. $W_i^n$'s are losslessly obtained at the decoder and then $U_i^n$'s are estimated. In a similar way LT can be extended to $N$ users.

We compare the performance of AF, SB and LT for $N = 2, 3$ in Fig 4. We see that AF performs well over a larger SNR region compared to the other two schemes as $N$ increases.

## IV. OPTIMAL POWER ALLOCATION

Till now we studied the performance of the three coding schemes for given average powers $P_1$ and $P_2$. The question arises if indeed it is optimal to expend the whole of $P_1$ and $P_2$ or one could do better by using lower average powers. Of course if the two users are independent of each other then we should expend full powers $P_1$ and $P_2$. In this section, using the performance obtained in previous sections we address this issue for the correlated case.

Without loss of generality we take the sources with unit variance (when they have different variances, by scaling we can reduce the problem to equal, unit variance case). Also, let MAC noise power $\sigma_N^2 = 1$. The objective is to find for the three schemes, the minimum distortion for the given power constraints $P_1$ and $P_2$ and also the optimal powers.

### A. Amplify and forward scheme

For the AF, we find powers $a^*$ and $b^*$ that minimize

$$\beta_1 D_1 + \beta_2 D_2 = \beta_1 \frac{b(1-\rho^2)+1}{a+b+2\sqrt{ab}\rho+1} + \beta_2 \frac{a(1-\rho^2)+1}{a+b+2\sqrt{ab}\rho+1} \tag{20}$$

subject to $a \leq P_1$ and $b \leq P_2$, where $\beta_1, \beta_2$ are given positive weights.

Unlike for a single user case, the optimal powers $a^*$ and $b^*$ may not be equal to $P_1$ and $P_2$ respectively. Therefore, we solve this optimization problem using the Kuhn-Tucker first order conditions ([4]). In the following we also study the qualitative behavior of the cost function to get more insight into the problem.

From Section III-D for symmetric case we know that $D$ strictly decreases as a function of $P$. Hence then for $\beta_1 = \beta_2$, $a^* = b^* = P$. In general, if we use powers $a$ and $b$, then

$$D_1 + D_2 > \frac{(a+b)(1-\rho^2)}{a+b+2\sqrt{ab}\rho} \geq \frac{1-\rho^2}{1+\rho} = 1-\rho.$$

The second inequality follows because $a + b \geq 2\sqrt{ab}$ and equality is achieved only if $a = b$. The first inequality will tend to equality if $a$ or $b \to \infty$. Thus for large $P_1$ and $P_2$, $a^* \approx b^* \approx min(P_1, P_2)$. The minimum $D_1 + D_2$ approaches the lower bound $1 - \rho$ as $P_1$ and $P_2 \to \infty$. However if we keep $a$ fixed and increase $b$ then the first lower bound starts increasing after sometimes. Thus if $P_1$ is fixed and $P_2$ is increased, $b^*$ will not keep increasing. We will elaborate on it below via explicit computations.

From (20) we find that $D_1$ is monotonically decreasing with $a$. Next fix $a$ and vary $b$. If $a$ and $b$ are large compared to 1 and if $a \ll b$ then $D_1 \approx 1 - \rho^2$. Thus $D_1$ is insensitive to $b$ if $1 \ll a \ll b$. This can be interpreted as follows. At high $b$ the noise can be neglected and when $1 \ll a \ll b$ at the decoder $U_2$ is available accurately and $Y$ will be close to $U_2$. Thus $E[U_1|U_2]$ will be close $E[U_1|Y]$ and the mean square error $E\left[(U_1 - \hat{U}_1)^2\right] \approx 1 - \rho^2$. By symmetry, same follows for $D_2$ when $1 \ll b \ll a$.

Next consider $D_1$ when $1 \ll b \ll a$. Then $D_1 \approx \frac{b(1-\rho^2)}{a}$. Here $D_1$ is sensitive to $b$ and increases with $b$ irrespective of $\rho$. This may be because the interference effect of $U_2$ dominates as $b$ increases.

Now we again consider the case where $P_1$ is fixed and $P_2$ is varying. The optimal values of $a^*, b^*$ and $D$ are provided in Table I for $\rho = 0.5$ via Kuhn-Tucker conditions. As we see, $b^*$ is less than $P_2$ for large $P_2$. This happens because when $a$ is fixed, $D_1 + D_2$, as a function of $b$, is strictly decreasing in the interval $[0, c)$ and strictly increasing in $(c, \infty)$ where $c = [((1+\rho^2) + \sqrt{(1+\rho^2)^2 + 4a\rho^2(1-\rho^2)[a(1-\rho^2)+2]})/2\sqrt{a}\rho(1-\rho^2)]^2$.

Thus $b = c$ is the unique global minimum of $D_1 + D_2$ for a given $a$. From Table I we observe that for $\rho = .5$ and $a = 10$, $c = 17.01$. Table I also shows that for $P_2 > P_1$, $a^* = P_1$.

## B. Separation based scheme

For the SB scheme, $X_1, X_2$ are independent zero mean Gaussian random variables. The capacity region for this case monotonically increases with $P_1, P_2$. Thus, the distortion $D_1 + D_2$ will be minimized at the powers $(a^*, b^*) = (P_1, P_2)$.

## C. Lapidoth-Tinguely Scheme

The optimization problem for the LT scheme is: Minimize

$$\beta_1 \frac{\sigma_1^2 2^{-2R_1}[1-\rho^2(1-2^{-2R_2})]}{1-\tilde{\rho}^2} + \beta_2 \frac{\sigma_2^2 2^{-2R_2}[1-\rho^2(1-2^{-2R_1})]}{1-\tilde{\rho}^2} \quad (21)$$

Subject to, $a \leq P_1$, $b \leq P_2$ and

$$R_1 \leq 0.5 \log\left(a + \frac{1}{1-\tilde{\rho}^2}\right), \ R_2 \leq 0.5 \log\left(b + \frac{1}{1-\tilde{\rho}^2}\right), \ R_1 + R_2 \leq 0.5 \log\left(\frac{1}{1-\tilde{\rho}^2} + \frac{a+b+2\sqrt{ab}\tilde{\rho}}{1-\tilde{\rho}^2}\right).$$

From the constraints we find that $R_1$, $R_2$ and $R_1 + R_2$ are monotonically increasing in $a, b$ and $\tilde{\rho}$. Hence these are maximized at $(a,b) = (P_1, P_2)$. Also the objective function is monotonically decreasing in $R_1$ and $R_2$ (the first derivative of the objective function is negative for all non-negative $(R_1, R_2)$ for all positive $\beta_1$, $\beta_2$). Hence the minimum distortion is achieved at powers $(P_1, P_2)$.

## V. SIDE INFORMATION

In a sensor network the cluster head, to which the neighbouring nodes may transmit their information via a MAC, may also be sensing. The sensed observations at the cluster head may be correlated to the data being transmitted by other nodes. This leads to the situation where the decoder has side information which can help reduce the transmitted data (see Theorem 1). The side information may also be available at the encoders because the sensor nodes transmitting their data can also listen to the data transmitted by other nodes to the cluster head. Thus in this section we compare the performance of the three schemes when there is side information at the encoders and/or the decoder.

Side information $Z_i$ is available at encoder $i$, $i = 1, 2$ and $Z$ is available at the decoder. One use of the side information $Z_i$ at the encoders is to increase the correlation between the sources. In our setup, then $W_i^n$ will be a function of $(U_i^n, Z_i^n)$. It is known that if random vectors $(V_{11}, ..., V_{1p})$ and $(V_{21}, ..., V_{2q})$ have a joint normal distribution then the functions $f(V_{11}, ..., V_{1p})$, $g(V_{21}, ..., V_{2q})$ having maximum correlations among themselves are linear (see [5]). This motivates us to take appropriate linear combinations $L_i = a_i U_i + b_i Z_i$, $i = 1, 2$ of $(U_i, Z_i)$ at encoder $i$ to generate $W_i$. When side information is available at the decoder it is used to estimate the sources and also it can reduce the rates through the channel (as can be seen from (1)). Using these ideas in the following we propose techniques to extend the three schemes to the case when there is side information at the encoders and the decoder. These techniques can be specialized to the case when there is encoder only or decoder only side information.

## A. AF with side information

Linear combination of the source outputs and side information $L_i = a_i U_i + b_i Z_i$, $i = 1, 2$ is amplified to $X_i$ to meet the power constraints and sent over the channel. These linear combinations can increase the correlation between the channel alphabets. However a distortion may be incurred in estimating the sources from $L_i$. Thus the optimal $a_i s$ and $b_i s$ selected may not give the highest possible correlation between $L_1$ and $L_2$. The decoder side information $Z$ is used to get better estimates of the sources. Hence we find the linear combinations, which minimize the sum of distortions. For this we consider the following optimization problem: find $(a_1, b_1, a_2, b_2)$ that

$$Minimize \ D(a_1, b_1, a_2, b_2) = E[(U_1 - \hat{U}_1)^2] + E[(U_2 - \hat{U}_2)^2] \quad (22)$$

subject to $E[X_i^2] \leq P_i$ where $\hat{U}_i = E[U_i|Y,Z]$, $i = 1, \ 2$.

## B. SB with side information

For a given $(L_1, L_2)$, we use the coding-decoding scheme described in Section III-B. The availability of the side information $Z$ at the decoder reduces the rates $(R_1, R_2)$ at which the vector quantizers operate. Side information $Z$ is also useful in estimating the sources. The linear combinations $L_1$ and $L_2$ are obtained which minimize (22) for this coding-decoding scheme.

## C. LT with side information

For a given $(L_1, L_2)$, we use the encoding-decoding scheme described in Section III-C by replacing $(U_1^n, U_2^n)$ by $(L_1^n, L_2^n)$. The rates at which the vector quantizers operate are a function of the side information at the decoder and are obtained from (1). This side information is also used in estimating $(U_1, U_2)$ from $(W_1, W_2)$. The linear combinations $L_1$ and $L_2$ are obtained which minimize (22) for this encoding-decoding scheme.

## D. Comparison of the schemes with side information

We provide the comparison of the three schemes for $U_1, U_2 \sim \mathcal{N}(0,1)$ and correlation $\rho$. Also for illustration purposes, we take the side information with a specific structure which seems natural in this set up. Let $Z_1 = s_1 U_2 + V_1$ and $Z_2 = s_2 U_1 + V_2$, where $V_1, V_2 \sim \mathcal{N}(0,1)$ and are independent of each other and independent of the sources, and $s_1$ and $s_2$ are constants that can be interpreted as the side-channel SNR. This can also be interpreted in the framework of co-operative communication where $Z_1$ and $Z_2$ represent the co-operation between the encoders. We also take $Z = (Z_1, Z_2)$.

For simplicity, we consider the symmetric case: $P_1 = P_2 = P$, $s_1 = s_2$, $D_1 = D_2$. The channel noise variance $\sigma_N^2 = 1$. Thus $P$ can be interpreted as the channel SNR. We obtain the optimal $a_i$, $b_i$ for the three schemes and compare the distortions as in Section III.

Consider the AF scheme. Fig. 5 gives the minimum sum of distortions achieved for $\rho = 0.1$ for different channel SNRs under various assumptions on the availability of the side information. It can be seen from the figure that decoder only side information is much more useful than encoder only side information. The reduction in distortion is directly proportional to the quality of the side information (i.e., the side channel SNR). The encoder only side information case shows marginal improvement over the no side information case. It is also found that when side information is provided at the decoder, providing it at the encoders also does not help much as seen in Fig. 5. Thus the critical information is the side information at the decoder. This is consistent with the findings in [8].

For SB and LT schemes also we find that decoder side information is the most useful one. It is also found that for the encoder only side information LT performs better than SB under all SNRs and all $\rho$. This conclusion is similar to the no side information case. AF performs better than both the schemes for low SNR's. These curves are not provided due to lack of space.

Decoder-only side information is used for estimation at the decoder and also for determining the communication rates at the encoders. Comparison of the three schemes for this case is shown in Fig. 6. It is seen that SB becomes better than LT when the quality of the side information provided at the decoder becomes better. In the symmetric case this cut-off (where SB takes over LT) side channel SNR is a function of $\rho$. Fig. 6 plots the sum of distortions vs the side channel SNR $s_1 = s_2 = s$ and gives the cut-off side channel SNR above which the SB is optimal when $\rho = 0.5$. This cut-off is valid for all channel SNR's. SB becomes optimal because the estimation efficiency and the communication rates can be improved in SB when the quality of the side information improves. The same effect is found when side information is fed to both encoders and the decoder (results not shown).

Overall, for the case of side information, we obtain the following conclusions. Decoder only side information is much more useful than encoder only side information. Encoder only side information helps marginally if at all. The reduction in distortion is proportional to the side information quality. Also, AF is optimal at low SNR with or without side information. However distortions in AF do not go to zero, when channel SNR is increased, with or without side information. But distortions for SB and LT do go to zero as channel SNR increases. Therefore, at high enough SNR AF is the worst. This is because of the interference between the two users. LT is always better than SB in the no side information case. But with side information SB is sometimes better than LT.

## VI. Transmission of correlated sources over orthogonal channels

For the orthogonal channel system $Y = (Y_1, Y_2)$ and the channel transition matrix is $p(y_1, y_2|x_1, x_2) = p(y_1|x_1)p(y_2|x_2)$. Then the conditions (1) become

$$I(U_1, Z_1; W_1|W_2, Z) < I(X_1; Y_1|W_2, Z) \leq I(X_1; Y_1), \tag{23}$$

$$I(U_2, Z_2; W_2|W_1, Z) < I(X_2; Y_2|W_1, Z) \leq I(X_2; Y_2), \tag{24}$$

$$I(U_1, U_2, Z_1, Z_2; W_1, W_2|Z) < I(X_1, X_2; Y_1, Y_2|Z) \leq I(X_1; Y_1) + I(X_2; Y_2). \tag{25}$$

Using Fano's inequality, for lossless transmission of discrete sources over discrete channels with side information, we can show that the outer bounds in (23)-(25) are in fact necessary and sufficient conditions. The outer bounds in (23)-(25) are satisfied with equality if the channel codewords $(X_1, X_2)$ are independent of each other. Also, the distribution of $(X_1, X_2)$ maximizing these bounds is not dependent on the distribution of $(U_1, U_2)$. Furthermore, the left side of the inequalities are simultaneously minimized when $W_1$ and $W_2$ are independent. Thus, the source coding of $(U_1, Z_1)$ and $(U_2, Z_2)$ can be done as in Slepian-Wolf coding but also taking into account the fact that the side information $Z$ is available at the decoder.

If we take $W_1 = U_1$ and $W_2 = U_2$ and the side information $(Z_1, Z_2, Z) \perp (U_1, U_2)$ ($X \perp Y$ denotes that $X$ is independent of $Y$) we can recover the conditions in [3].

### A. Gaussian sources over orthogonal Gaussian channels

Now we consider the transmission of jointly Gaussian sources over orthogonal Gaussian channels. Initially it is assumed that there is no side information $Z_1, Z_2, Z$.

Let $(U_1, U_2)$ be zero mean jointly Gaussian random variables with variances $\sigma_1^2$ and $\sigma_2^2$ respectively and correlation $\rho$. The channel outputs $Y_i = X_i + N_i, i = 1, 2$ where $N_i$ is Gaussian with zero mean and variance $\sigma_{N_i}^2$. Also $N_1$ and $N_2$ are independent of each other and also of $(U_1, U_2)$.

In this scenario, the right side of the inequalities in (23)-(25) are maximized by taking $X_i \sim \mathcal{N}(0, P_i)$, $i = 1, 2$, independent of each other where $P_i$ is the average transmit power constraint on user $i$. Then $I(X_i, Y_i) = 0.5 log(1 + P_i/\sigma_{N_i}^2)$, $i = 1, 2$.

We can easily specialize the above results to a TDMA, FDMA or CDMA based transmission scheme.

Thus, now as commented in the last section, for the orthogonal channels (without side information) an optimal scheme would be to use Slepian-Wolf source coding and then channel coding each source optimally as in a point to point communication. In fact for the Gaussian sources with orthogonal Gaussian channels and no side information, vector quantization of sources followed by Slepian-Wolf coding is an optimal source coding scheme ([24]) for two user case. This followed by $iid$ Gaussian coded sequences with mean 0 and variance $P_i$, $i = 1, 2$ will provide the overall optimal scheme for this system. The optimality is due to the fact that separation holds for this system ([26]). This is the SB scheme we have been studying in this paper.

In the following we compare the performance of the Amplify and Forward (AF) scheme, which makes the sensor node design simple, with the SB scheme. Unlike in the GMAC there is no interference between the two users when orthogonal channels are used. Therefore, in this case we expect AF to perform quite well even at high SNR.

## VII. AMPLIFY AND FORWARD OVER ORTHOGONAL CHANNELS

In this section we study the performance of AF in transmitting correlated Gaussian sources over orthogonal channels. For minimum mean square distortion, the decoder calculates conditional expectations $E[U_1|Y_1, Y_2]$ and $E[U_2|Y_1, Y_2]$. The distortions incurred are given by the respective conditional variances. The minimum distortions $(D_1, D_2)$ are

$$D_1 = \frac{(\sigma_1 \sigma_{N_1})^2 \left[P_2(1-\rho^2) + \sigma_{N_2}^2\right]}{P_1 P_2(1-\rho^2) + \sigma_{N_2}^2 P_1 + \sigma_{N_1}^2 P_2 + \sigma_{N_1}^2 \sigma_{N_2}^2}, \quad D_2 = \frac{(\sigma_2 \sigma_{N_2})^2 \left[P_1(1-\rho^2) + \sigma_{N_1}^2\right]}{P_1 P_2(1-\rho^2) + \sigma_{N_2}^2 P_1 + \sigma_{N_1}^2 P_2 + \sigma_{N_1}^2 \sigma_{N_2}^2}. \quad (26)$$

From (26) we see that as $P_1, P_2 \to \infty$ the distortions $D_1, D_2$ tend to zero. We also see that $D_1$ and $D_2$ are minimum when the average powers used are $P_1$ and $P_2$. These conclusions are in contrast to the case of a GMAC where the distortion for the AF does not approach zero as $P_1, P_2 \to \infty$ and the optimal powers needed may not be $P_1$ and $P_2$.

### A. Comparison of AF with SB

For comparing the performance of the two schemes we consider the symmetric case where $P_1 = P_2 = P, \sigma_1^2 = \sigma_2^2 = \sigma^2, D_1 = D_2 = D, \sigma_{N_1}^2 = \sigma_{N_2}^2 = \sigma_N^2$.

For the present system, the minimum distortions achieved in SB and AF are denoted by $D(SB)$ and $D(AF)$ respectively. $\sigma^2$ is taken to be unity without loss of generality. We denote $P/\sigma_N^2$ by $S$. Then

$D(SB)$ and $D(AF)$ are

$$D(SB) = \sqrt{\frac{1-\rho^2}{(1+S)^2} + \frac{\rho^2}{(1+S)^4}}, \quad (27)$$

$$D(AF) = \frac{S(1-\rho^2)+1}{1+2S+S^2(1-\rho^2)}. \quad (28)$$

We see from the above equations that when $\rho = 0$, $D(SB) = D(AF) = 1/(1+S)$. At high $S$, $D(AF) \approx 1/S$ and $D(SB) \approx \sqrt{1-\rho^2}/S$. Eventually both $D(SB)$ and $D(AF)$ tend to zero as $S \to \infty$. When $S \to 0$ both $D(SB)$ and $D(AF)$ go to $\sigma^2$.

By squaring the equations (27) and (28) we can show that $D(AF) \geq D(SB)$ for all $S$. But in the following we show that $D(AF)$ is often close to $D(SB)$.

We note that both $D(AF)$ and $D(SB)$ are lower bounded by $S(1-\rho^2)/(1+S)^2$. We denote this by $D(LB)$. $D(AF)$ is also upper bounded by $(1+S)/[1+S(1-\rho^2)]^2$, which we denote by $D(UB)$. Then,

$$D(AF) - D(SB) \leq D(UB) - D(LB) \leq \frac{\rho^2}{S}\left[\frac{1}{(1-\rho^2)^2} + \frac{1}{(1-\rho^2)} + 1\right]. \quad (29)$$

Thus, we see that for large $S$ the difference is small and tends to 0 as $S \to \infty$.

Again, from (28), $D(AF) \leq S(1-\rho^2) + 1$. This gives,

$$D(AF) - D(SB) \leq \frac{S(2+S)(1+S(1-\rho^2))}{(1+S)^2} \leq S + \frac{S}{1+S}. \quad (30)$$

The right side is small when $S$ is small. The difference is $O(S)$ as $S \to 0$. $D(AF) - D(SB)$ is less than the minimum of the right side in (29) and (30). Also,

$$D(AF) - D(SB) \leq D(AF) - D(LB) \leq \frac{S^2\rho^2[1+S(1-\rho^2)]}{(1+S)^2(1+2S+S^2(1-\rho^2))}$$
$$\leq \frac{\rho^2[1+S(1-\rho^2)]}{(1+2S+S^2(1-\rho^2))} \leq \frac{\rho^2}{1+S(1-\rho^2)} \leq \rho^2. \quad (31)$$

From (29), (30) and (31) we conclude that $D(AF) - D(SB)$ is small when $S$ is small or large or whenever $\rho$ is small.

$D(AF)$ and $D(SB)$ are plotted for $\rho$=0.3 and 0.7 using exact computations in Figs 7 and 8. These figures confirm the theoretical findings.

*B. Side information*

Let us consider the case when side information $Z_i$ is available at encoder $i$, $i = 1, 2$ and $Z$ is available at the decoder. One use of the side information $Z_i$ at the encoders is to increase the correlation between the sources. Then while using SB, the encoding rates at the two sources can be decreased. This can be optimally done (see [5]), if we take appropriate linear combination of $(U_i, Z_i)$ at encoder $i$ and use SB, as done in Section V.

We provide the comparison of AF with SB for $U_1, U_2 \sim \mathcal{N}(0,1)$. Also we take the side information which has the same structure as in Section V-D.

We have compared AF and SB with different $\rho$ and $s_1, s_2$ by explicitly computing the minimum $(D_1 + D_2)/2$ achievable. We take $P_1 = P_2$. Due to lack of space we provide only one case in Fig.9 where $s_1 = s_2 = 0.5$ and $\rho = 0.4$. From the figure one sees that without side information, the performance of AF and SB is very close for different SNRs. The difference in their performance increases with side information for moderate values of SNR because the effect of the side information is to effectively increase the correlation between the sources. Even for these cases at low and high SNRs the performance of AF is close to that of SB.

For the symmetric case discussed here, for SB, encoder-only side information reduces the distortion marginally. This happens because a distortion is incurred for $(U_1, U_2)$ while making the linear combinations $(L_1, L_2)$. For the AF we actually see no improvement and the optimal linear combination has $b_1 = b_2 = 0$. For decoder-only side information the performance is improved for both AF and SB as the side information can be used to obtain better estimates of $(U_1, U_2)$. Adding encoder side information further improves the performance only marginally for SB; the AF performance is not improved.

## VIII. CONCLUSIONS AND FUTURE DIRECTIONS

We have specialized a general result discussed in [20] to the Gaussian sources and the Gaussian MAC. Next, we analyze three joint source-channel coding schemes available in the literature and compare their distortion performance. We prove that the amplify and forward scheme (AF) is sub-optimal at high SNR's. Also we show that another scheme in [14] is asymptotically optimal as the SNR increases. Cases with side information are also addressed. We also provide optimum power allocation policies for the three schemes. Next we study orthogonal Gaussian channels. We identify an optimal joint source-channel coding scheme. Then we show that unlike the GMAC case, in orthogonal case the AF can often be close to the optimal scheme even at high SNR.

In future, it will be important to study the optimal schemes for moderate values of SNR. Also, of course one should obtain optimal schemes for transmission of non Gaussian sources over a GMAC. A good scheme has been provided in [19].

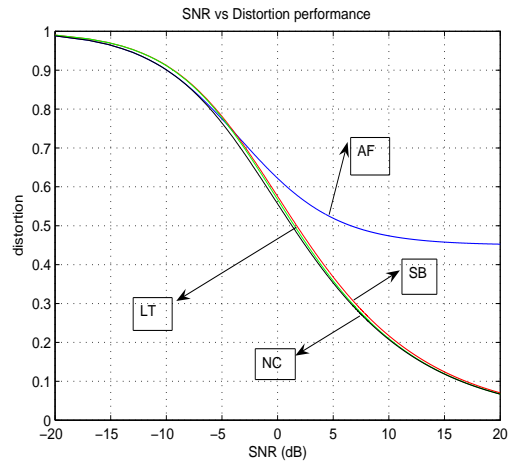

Fig. 1. Comparison of performance of the three schemes: SNR vs distortion for $\rho$=0.1.

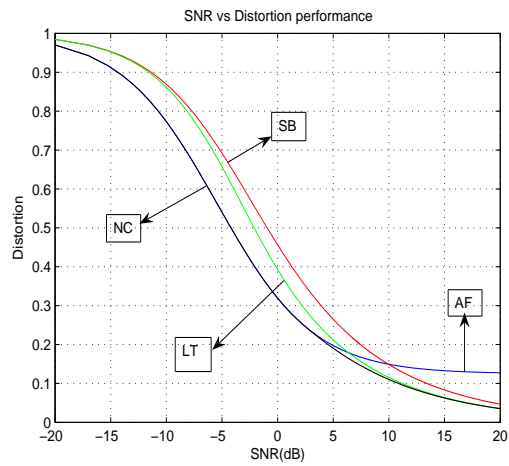

Fig. 2. Comparison of performance of the three schemes: SNR vs distortion for $\rho$=0.75.

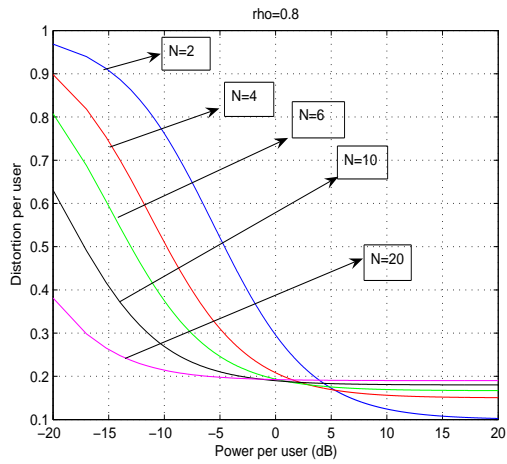

Fig. 3. Distortion for AF for different number of users, $\rho = 0.8$.

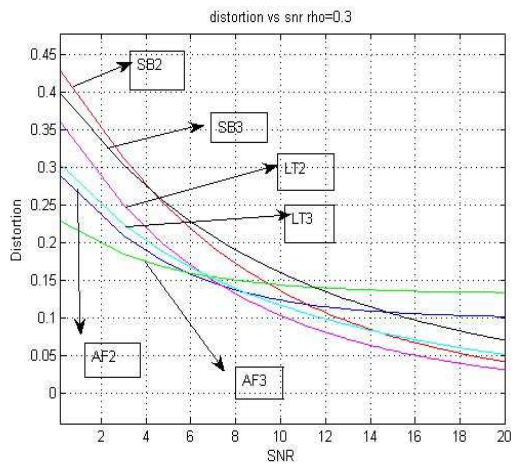

Fig. 4. Comparison of the three schemes for N=2, 3, $\rho = 0.8$.

TABLE I

OPTIMAL POWER ALLOCATION FOR AF FOR THE ASYMMETRIC CASE

| $P_1$ | $P_2$ | $a^*$ | $b^*$ | $D_{min}$ |
|---|---|---|---|---|
| 10 | 1 | 10 | 1 | 0.6760 |
| 10 | 5 | 10 | 5 | 0.5743 |
| 10 | 20 | 10 | 17.01 | 0.5422 |
| 10 | 50 | 10 | 17.01 | 0.5422 |

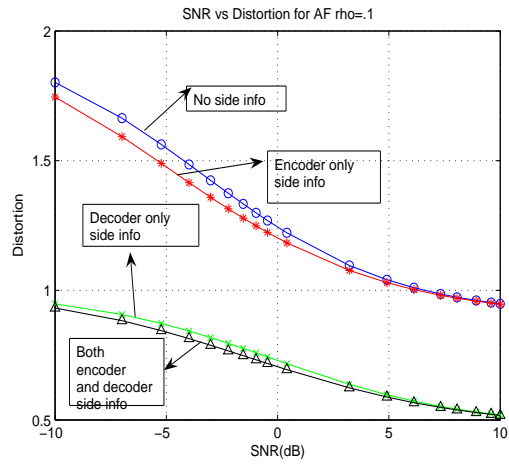

Fig. 5. SNR vs distortion performance for AF with side information, $s_1 = s_2 = 1$, $\rho = .1$.

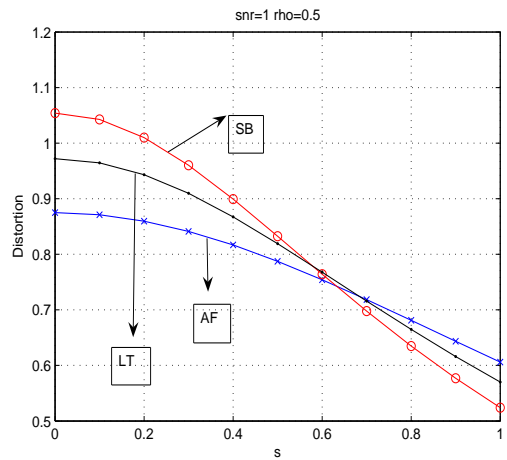

Fig. 6. Comparison of three schemes with decoder only side information: Distortion vs side information power, SNR=0dB, $\rho = .5$.

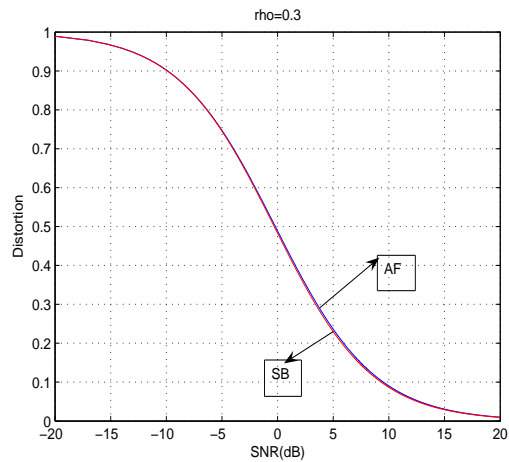

Fig. 7. Orthogonal channels: SNR vs distortion performance for SB vs AF $\rho = .3$.

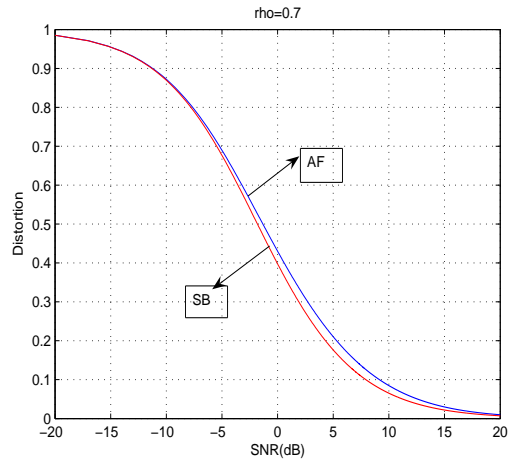

Fig. 8. Orthogonal channels: SNR vs distortion performance for SB vs AF $\rho = .7$.

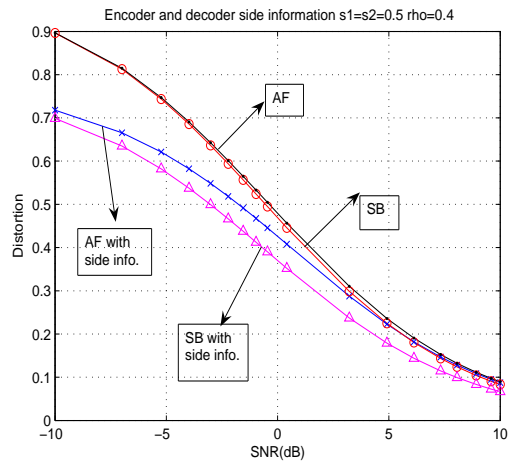

Fig. 9. Orthogonal channels: AF and SB with both encoder and decoder side information, $\rho = .4$.